\documentclass[review]{elsarticle}

\usepackage{color}
\usepackage{amssymb}
\usepackage{graphicx}
\usepackage{amsmath}
\usepackage{lineno,hyperref}
\modulolinenumbers[5]

\journal{Combustion and Flame}

\biboptions{longnamesfirst,sort&compress}
\begin{document}

\begin{frontmatter}
\title{Kinetic modeling of detonation and effects of negative temperature coefficient}

\author[BUAA]{Yudong Zhang}
\author[NLST,CAPT]{Aiguo Xu\corref{cor1}}
\author[CAPT]{Guangcai Zhang}
\author[BUAA]{Chengmin Zhu\corref{cor2}}
\author[CUMT]{Chuandong Lin}

\cortext[cor1]{Corresponding author. E-mail address: Xu\_Aiguo@iapcm.ac.cn}
\cortext[cor2]{Corresponding author. E-mail address: cmzhu@buaa.edu.cn}

\address[BUAA]{School of Astronautics, Beijing University of Aeronautics and Astronautics, Beijing, 100191, China}
\address[NLST]{Laboratory of Computational Physics, Institute of Applied Physics and Computational Mathematics, Beijing, 100088, China}
\address[CAPT]{Center for Applied Physics and Technology, MOE Key Center for High Energy Density Physics Simulations, College of Engineering, Peking University, Beijing 100871, China}
\address[CUMT]{State Key Laboratory for GeoMechanics and Deep Underground Engineering,
China University of Mining and Technology, Beijing 100083, China}

\begin{abstract}
The kinetic modeling and simulation of reactive flows, especially for those with detonation, are further investigated. From the theoretical side, a new set of hydrodynamic equations are deduced, where the viscous stress tensor and heat flux are replaced by two non-equilibrium quantities that have been defined in our previous work. The two non-equilibrium quantities are referred to as Non-Organized Momentum Flux (NOMF) and Non-Organized Energy Flux (NOEF), respectively, here. The numerical results of
viscous stress (heat flux) have a good agreement with those of
 NOMF (NOEF) near equilibrium state. Around sharp interfaces, the values of NOMF (NOEF) deviate reasonably from those of viscous stress (heat flux).
  Based on this hydrodynamic model, the relations between the two non-equilibrium quantities and entropy productions are established. Based on the discrete Boltzmann model, four kinds of detonation phenomena with different reaction rates, including Negative Temperature Coefficient (NTC) regime, are simulated and investigated. The differences of the four kinds of detonations are studied from three aspects: hydrodynamic quantities, non-equilibrium quantities and entropy productions. It is found that, the effects of NTC on hydrodynamic quantities are to lower the von-Neumann peaks of density, pressure, and velocity, to broaden the reaction zone, and to subdue the chemical reaction. It may also vanish the peak of temperature. Consequently, the effects of NTC are to widen the non-equilibrium regions and reduce the amplitude of the non-equilibrium effects in the reaction zone. Besides, it is also found that the (local) entropy production has three sources: the chemical reaction, NOEF and NOMF. As for the global entropy production in the system, the portion caused by reaction is much larger than the other two, and the portion caused by NOMF is larger than that by NOEF. Furthermore, the effect of  NTC is to widen the region with entropy production caused by reaction and lower the global entropy productions caused by reaction, NOMF and NOEF, which means that NTC drives detonation closer to an isentropic process.
\end{abstract}


\end{frontmatter}

%\linenumbers

\section{Introduction}

Detonation is a special case of combustion which is the major energy
conversion process and plays a dominant role in
the transportation and power generation. It is a kind of chemical reaction phenomenon accompanied with violent energy release \cite{Mader-book,Theory-Detonation-book,Jiang-AM2012,Shepherd-PCI2009}. The system with detonation can generally be regarded as a kind of chemical reactive flow. The controlled detonation has long been extensively used in various engineering problems. Typical examples are referred to Pulse Detonation Engine \cite{PDE}, Rotating Detonation Engine \cite{RDE1,RDE2}, Oblique Detonation ramjet-in-Tube \cite{ODRT}, etc.

A detonation process may involve many species of reactants and a large number of reactions. For example, the CH$_4$/air detonation involves $53$ kinds of species and $325$ reactions \cite{CH4-detail,ChenZ2016} and n-heptane/air includes $2540$ reversible elementary reactions among 556 species \cite{NTC1}. The reaction rate generally varies with the specific reaction. For a practical detonation, the varieties of reactant species, shock strength, local temperature, specific volume, premixing homogeneity may guide the reactions into different chains. Consequently, the final detonation process may show different mechanical and thermodynamical behaviours according to the specific conditions. Therefore, the global reaction rate may show non-monotonic dependence on the temperature, even though the reaction rate shows exponential dependence on temperature in common cases, just like what the Arrhenius model describes. In fact, the phenomena related to Negative Temperature Coefficient (NTC) have been observed \cite{Chemicalbook,NTC1,NTC2,NTC3}. Physically, the occurrence of NTC may lead to significant different detonation behaviours. To the authors' knowledge, however, its possible effects have not obtained careful investigations.

It has been realized that various non-equilibrium behaviours extensively exist in the combustion and detonation phenomena \cite{Ju-Review}. But those complicated behaviours and their possible effects have far from been well studied.
It has also been well known that the traditional hydrodynamic modeling based on Euler or Navier-Stokes (NS) equations is not enough to describe such complicated non-equilibrium behaviours. The spatial-temporal scales that those complicated non-equilibrium behaviours make effects are much larger than those that the molecular dynamics can access. Under such cases, to investigate the possible effects of the non-equilibrium behaviours, a kinetic model based on the Boltzmann equation becomes preferable.

As a special discretization of the Boltzmann equation, the Lattice Boltzmann Method (LBM) \cite{Succi-Book,Wu2015A,Liu2015Lattice,Swift1996Lattice,Wagner1997Breakdown} has long been
attempted to simulate combustion phenomena \cite{Succi1997,Filippova2000_2,Yu2002,Yamamoto2005,Lee2006,Chiavazzo2010,Chen2011,Chen2012}. The first work was given by Succi, et al \cite{Succi1997} in 1997.
In those previous studies, the LBM works as a kind of alternative numerical scheme. The combustion systems are described by some kinds of hydrodynamic models.

To extend the LBM to model and simulate the detonation phenomena with complicated non-equilibrium behaviours, at least two technical bottlenecks must be broken through. The first one is to extend its application range to the cases where the Mach number is larger than 1. The second is that the improved model must be some kinetic model which not only can recover, in the hydrodynamic limit, but also is beyond the traditional NS model.
As was shown in recent years, one solution to the first bottleneck is to come back to the Finite-Difference(FD) LBM \cite{FOP2012-Review}. In the FD-LBM,  the discretization of the particle velocity space is independent of the discretizations of the space and time. This independence, together with the flexibilities in choosing FD scheme and in discretizing the particle velocity space, makes it easier for the numerical system to satisfy the von-Neumann stability condition in the cases with high Mach number compressible flows.
The solution to the second bottleneck sees also significant progress in recent years. The LBM has been extended to investigate
various non-equilibrium behaviours in complex flows \cite{Succi-DBM2015,Montessori2015Lattice,FOP2012-Review,APC2015-Review}.
Via such a modeling some new physical insights into the complex flows have been obtained. The observations have also been promoting the development of related methodology. For example, the strength of the non-equilibrium increases in the spinodal decomposition stage and decreases in the domain growth stage. Consequently, it can work as a kind of physical criteria to discriminate the two stages \cite{SoftMatter2015}. Different kinds of interfaces, such as material interface and mechanical interface, compressive wave and rarefactive wave, show different specific non-equilibrium properties. Consequently, the Thermodynamic Non-Equilibrium(TNE) can be used to distinguish  various interfaces \cite{PRE2014}. The TNE behaviours in complex flows have also been used in interface-tracking scheme designs \cite{RTI-arXiv2015}. It has been found that the viscosity (heat conductivity) decreases the local TNE but increases global TNE around the detonation wave \cite{PRE2015-Combustion}.
Such an extended lattice Boltzmann kinetic model or discrete Boltzmann model (DBM) should follow more strictly some necessary kinetic moment relations of the local equilibrium distribution function $f^{eq}$. In a recent study a double-distribution function DBM was proposed, where one distribution function is used to describe the reactant, the other distribution function is used to describe the reaction product \cite{LinCNF2015}. This DBM corresponds to the so-called ``two-fluid'' hydrodynamic model for combustion.

Entropy production is a highly concerned quantity in both physics and engineering studies. From the physics side, it is helpful for understanding the complex non-equilibrium behaviours. From the engineering side, a process with lower entropy production may have a higher energy transformation efficiency.

The objective of the present work is two-fold. Firstly, we further develop the DBM to investigate various non-equilibrium behaviours in combustion, especially in detonation phenomena, aiming to establish a relation between the TNE and the entropy production. Secondly, via a newly composed reaction function, we investigate the possible influences of the NTC on the behaviour of detonation in hydrodynamic quantities, TNE and entropy productions.

The organization of the present paper is as below. In section \ref{Sec.Models} we briefly review the DBM and the newly composed reaction rate function, present the new form of fluid hydrodynamic equations, and establish the relations between TNE and entropy production. In section \ref{Sec.Verfica} we validate the model by simulating two classical detonation benchmarks. In section \ref{Sec.Simulate}, we simulate four kinds of detonations with different temperature dependent reaction rates, analyze the differences of the four cases, and summarize the effects of NTC on detonation. Section \ref{Sec.Conclu} concludes the present paper.

\section{Models and methods}
\label{Sec.Models}

\subsection{Kinetic and Hydrodynamic models, non-equilibrium effects and entropy production}

The kinetic model based on Boltzmann equation with chemical reaction has a form as follows:
\begin{equation}
\frac{\partial f}{\partial t}+\mathbf{v}\cdot \nabla f=\Omega+C \mathtt{,} \label{Eq9_Boltzmann equation}
\end{equation}
where $f$ indicates the distribution function of particle velocity, $\mathbf{v}$ indicates the particle velocity, $t$ is temporal coordinate, $\Omega$ and $C$ are the collision term and chemical reaction term, respectively. The equilibrium distribution function of velocity particle in Eq. (\ref{Eq9_Boltzmann equation}) reads:
\begin{equation}
f^{eq}=\rho(\frac{1}{2\pi T})^{D/2}(\frac{1}{2n\pi T})^{1/2}\mathrm{exp}[-\frac{(\mathbf{v-u})^2}{2T}-\frac{\eta^2}{2nT}] \mathtt{.} \label{Eq10_equilibrium distribution function}
\end{equation}
where $D$ indicates spatial dimension, $\mathbf{v}$ and $\mathbf{u}$ are particle velocity and hydrodynamic velocity, respectively. $\eta$ is a free parameter introduced to describe the $n$ extra degrees of freedom corresponding to molecular rotation and/or vibration \cite{GanEPL2013}.
The central moment $\mathbf{M}_2^*$, $\mathbf{M}_{3,1}^*$ and thermodynamic non-equilibrium quantities $\mathbf{\Delta}_2^*$, $\mathbf{\Delta}_{3,1}^*$ are defined as \cite{Xureview2014}
\begin{equation}
\mathbf{M}_2^*(f)=\int\int f\mathbf{(v-u)(v-u)}d\mathbf{v}d\eta  \mathtt{,}  \label{Eq11_M2*}
\end{equation}
\begin{equation}
\mathbf{M}_{3,1}^*(f)=\int\int f\mathbf{(v-u)\cdot(v-u)(v-u)}d\mathbf{v}d\eta   \mathtt{,}  \label{Eq12_M31*}
\end{equation}
\begin{equation}
\mathbf{\Delta}_2^*=\mathbf{M}_2^*(f)-\mathbf{M}_2^*(f^{eq})  \mathtt{,}  \label{Eq13_Delta2*}
\end{equation}
\begin{equation}
\mathbf{\Delta}_{3,1}^*=\mathbf{M}_{3,1}^*(f)-\mathbf{M}_{3,1}^*(f^{eq}) \mathtt{.}  \label{Eq14_Delta31*}
\end{equation}

Taking the velocity moment $\int\int d\mathbf{v}d\eta$ of the Eq. (\ref{Eq9_Boltzmann equation}) gives
the continuity equation:
 \begin{equation}
\frac{\partial\rho}{\partial t}+\bigtriangledown\cdot(\rho\mathbf{u})=0    \mathtt{.}  \label{Eq15_continuity equation}
\end{equation}
Taking the velocity moment $\int\int\mathbf{v} d\mathbf{v}d\eta$ of the Eq. (\ref{Eq9_Boltzmann equation}) gives the momentum conservation equation:
\begin{equation}
\frac{\partial\rho\mathbf{u}}{\partial t}+\bigtriangledown\cdot(\rho\mathbf{uu}+P\mathbf{I}+\mathbf{\Delta}_2^*)=0    \mathtt{.}  \label{Eq16_momentum conservation equation}
\end{equation}
Taking the velocity moment $\int\int({\frac{\mathbf{v}^2}{2}+\frac{\eta^2}{2}})d\mathbf{v}d\eta$ of the Eq. (\ref{Eq9_Boltzmann equation}) gives the energy conservation equation:
\begin{equation}
\frac{\partial\rho(e+\frac{\mathbf{u}^2}{2})}{\partial t}+\bigtriangledown\cdot[\rho\mathbf{u}(e+T+\frac{\mathbf{u}^2}{2})+\mathbf{\Delta}_2^*\cdot\mathbf{u}+\mathbf{\Delta}_{3,1}^*]=\rho QF(\lambda)    \mathtt{.}  \label{Eq17_energy conservation equation}
\end{equation}
Comparing with NS equations as shown in Eqs. (\ref{NS1})-(\ref{NS3}) \cite{Minoru2003Two}, we can conclude that $\mathbf{\Delta}_2^*$ corresponds to the viscous stress tensor $\mathbf{\Pi}$, and  $\mathbf{\Delta}_{3,1}^*$ corresponds to the heat flux $\mathbf{j}_q$. Here we refer  $\mathbf{\Delta}_2^*$ to as Non-Organized Momentum Fluxes (NOMF), and refer $\mathbf{\Delta}_{3,1}^*$ to as Non-Organized Energy Fluxes (NOEF). The hydrodynamic models, Eqs. (\ref{Eq15_continuity equation})-(\ref{Eq17_energy conservation equation}), are derived from the Boltzmann equation, Eq.(\ref{Eq9_Boltzmann equation}), with the complete distribution function $f$. While, the common NS models , Eqs. (\ref{NS1})-(\ref{NS2}), are derived from the Boltzmann equation, with the approximation, $f \approx f^{0} + f^{1}$, where the Knudsen number has been absorbed in $f^{1}$, $f^{0}$ is the local thermodynamic equilibrium distribution function and $f^{1}$ is the first-order deviation of $f$ from $f^{0}$. Consequently, the quantity, $\mathbf{\Delta}_2^*$ ($\mathbf{\Delta}_{3,1}^*$), contains more information than $\mathbf{\Pi}$ ($\mathbf{j}_q$) in NS equations. The numerical comparisons of $\mathbf{\Pi}$ and $\mathbf{\Delta}_2^*$, $\mathbf{j}_q$ and $\mathbf{\Delta}_{3,1}^*$  are made in section \ref{Sec.non-equil}.

\begin{subequations}
\begin{equation}
\frac{\partial \rho}{\partial t} + \nabla\cdot(\rho\mathbf{u}) = 0 \mathtt{,} \label{NS1}
\end{equation}
\begin{equation}
\frac{{\partial \left( {\rho {\bf{u}}} \right)}}{{\partial t}} + \nabla  \cdot \left[ {\rho {\bf{uu}} + P {\bf{I}} + {\bf {\Pi} }} \right] = 0  \mathtt{,} \label{NS2}
\end{equation}
\begin{equation}
\frac{{\partial \left[ {\rho (e + \frac{{{{\bf{u}}^2}}}{2})} \right]}}{{\partial t}} + \nabla  \cdot \left[ {\rho {\bf{u}}\left( {e + T + \frac{{{{\bf{u}}^2}}}{2}} \right) +  {{\bf{j_q}} + {\bf {\Pi} } \cdot {\bf{u}}} } \right] = 0  \mathtt{,} \label{NS3}
\end{equation}
\end{subequations}
with
\begin{equation}
  {\bf {\Pi} } = {\mu \left( {\nabla {\bf{u}} + {{(\nabla {\bf{u}})}^T} - \frac{2}{{D + n}}{\bf{I}}\nabla  \cdot {\bf{u}}} \right)}  \mathtt{,} \label{Pi}
\end{equation}
\begin{equation}
{\bf{j_q}} = {\kappa '\nabla T} \mathtt{.} \label{j_q}
\end{equation}
where $\mu$ and $\kappa '$ are viscosity coefficient and heat conductivity, respectively.

Following the way of defining entropy equilibrium equation in the non-equilibrium thermodynamics \cite{Non-thermobook}, replacing $\mathbf{\Pi}$ and $\mathbf{j}_q$ with $\mathbf{\Delta}_2^*$ and $\mathbf{\Delta}_{3,1}^*$, respectively, we get a new entropy equilibrium equation as follows:
\begin{equation}
\frac{\partial s}{\partial t}=-\bigtriangledown\cdot(s\mathbf{u}+\frac{1}{T}\mathbf{\Delta}_{3,1}^*)
+\mathbf{\Delta}_{3,1}^*\cdot\bigtriangledown(\frac{1}{T})-\frac{1}{T}\mathbf{\Delta}_2^*:\bigtriangledown(\mathbf{u})
+\rho\frac{Q}{T}F(\lambda)    \mathtt{.}  \label{Eq18_entropy formation1}
\end{equation}
where $F(\lambda)$ indicates the reaction rate function that will be defined later.
In fact, the $\frac{\partial s}{\partial t}$ can be written as
\begin{equation}
\frac{\partial s}{\partial t}=-\bigtriangledown\cdot J_s
+\sigma   \mathtt{.}  \label{Eq19_entropy formation2}
\end{equation}
The two terms on the right hand side of Eq. (\ref{Eq19_entropy formation2}) are called entropy flux and entropy production, respectively. Comparing Eqs. (\ref{Eq18_entropy formation1}) and (\ref{Eq19_entropy formation2}) gives
\begin{equation}
J_s=s\mathbf{u}+\frac{1}{T}\mathbf{\Delta}_{3,1}^* \mathtt{,} \label{Eq20_entropy flux}
\end{equation}
\begin{equation}
\sigma=\mathbf{\Delta}_{3,1}^*\cdot\bigtriangledown(\frac{1}{T})-\frac{1}{T}\mathbf{\Delta}_2^*:\bigtriangledown(\mathbf{u})
+\rho\frac{Q}{T}F(\lambda)   \mathtt{.}  \label{Eq21_entropy production}
\end{equation}

From Eq. (\ref{Eq21_entropy production}) it is clear that there are three source terms in entropy production. The first term is caused by NOEF, the second is caused by NOMF, and the third is caused by chemical reaction. In the following sections, we will compare the three portions of entropy production during detonation under different temperature-dependent reaction rates.
\subsection{Discrete Boltzmann model}

In 2013, an uniform scheme for composing single-relaxation-time and multiple-relaxation-time DBM was proposed \cite{GanEPL2013}. In this scheme the discrete local equilibrium distribution function $f_{i}^{eq}$ is inversely calculated from the kinetic moment relations it satisfies. Our studies in this work are based on a hybrid model coupled by a DBM for high speed compressible flows in Ref. \cite{GanEPL2013} and a phenomenological reaction rate function \cite{ZhangCMP2015}.

Mathematically, the discrete model equations read
\begin{subequations}
\begin{equation}
\frac{\partial f_i}{\partial t}+\mathbf{v}_{i} \cdot \nabla f_i=-\frac{1}{\tau}(f_i-f^{*eq}_i) \mathtt{,} \label{Eq1}
\end{equation}
\begin{equation}
\frac{d\lambda}{dt}=
\left\{
\begin{array}{l}
k(1-\lambda)\lambda ,{\kern 12pt}  T \ge {T_{th}}{\kern 4pt} \mathrm{and} {\kern 4pt} 0 \le \lambda  \le 1  \\
0\tt{,}{\kern 56pt} \mathrm{else} \mathtt{.}\label{Eq2}
\end{array}
\right.
\end{equation}
\end{subequations}
where subscript $i$ indicates the $i$-th particle velocity, $\tau$ is the relaxation time, $t$ and $r_\alpha$ are temporal and spatial coordinate, respectively. $\lambda$ is a parameter indicating the chemical reaction process, $k$ indicates the reaction rate constant, $T_{th}$ is temperature threshold for chemical reaction. $f_i$ indicates the distribution function of particle velocity and $f_i^{*eq}$ indicates the local equilibrium distribution function containing the effect of chemical reaction, which reads
\begin{equation}
f_i^{*eq}=f_i^{eq}(\rho,\mathbf{u},e^*)=f_i^{eq}(\rho,\mathbf{u},e+\tau QF(\lambda))\mathtt{,} \label{Eq4}
\end{equation}
with
\begin{equation}
F(\lambda)=\frac{d\lambda}{dt} \mathtt{.} \label{Eq5}
\end{equation}
where $\rho, \mathbf{u}, e$ are density, velocity, and internal energy, respectively. $Q$ is the amount of heat released by the chemical reactant per unit mass.

Here we consider the case where the time scale of thermodynamic  relaxation is much smaller than that of chemical reaction. So, the effect of chemical reaction is dynamically considered in the calculation of equilibrium distribution function. The discrete velocity model adopted here (D2V16) and the method to solve $f_i^{eq}$ are the same as in Ref. \cite{GanEPL2013}.

%%\begin{figure}[tbp]
%%\center\includegraphics*
%%[bbllx=0pt,bblly=0pt,bburx=240pt,bbury=240pt,angle=0,width=0.5\textwidth]{Fig1.eps}
%%\caption{Schematic of the multiple time scales in the evolution of physical system with chemical reaction.}
%%\label{Fig1}
%%\end{figure}

%%\begin{figure}[h]
%%\centerline{\psfig{file=fig1.,width=2in}}
%%\vspace*{8pt}
%%\caption{Sketch of the Discrete velocity model.}\label{fig1}
%%\end{figure}

\begin{figure}[tbp]
\center\includegraphics*
[bbllx=0pt,bblly=0pt,bburx=410pt,bbury=320pt,angle=0,width=0.7\textwidth]{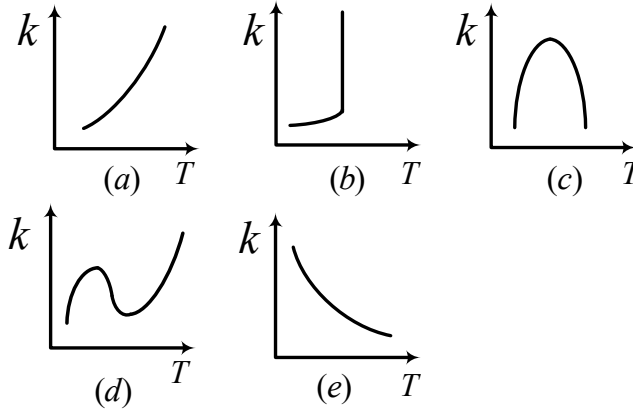}
\caption{The non-monotonic dependence of reaction rate constant ($k$) on temperature ($T$).}\label{fig2}
%%\caption{The influence of temperature on $k$. (a) The most commonly conditions that $k$ increase with temperature. (b) The explode reactions. (c) The enzymic catalytic reactions. (d) Reactions such as the oxidation of large hydrocarbons. (e) Reactions such as $2NO + O_2 \rightarrow2NO_2$.}\label{fig2}
\end{figure}

The impact of temperature on reaction rate is described by the dependence of $k$ on the temperature. In fact, the dependence is non-monotonic \cite{Chemicalbook} as shown in Fig. \ref{fig2}. Figure \ref{fig2}(a) is for the most common case that $k$ increases with temperature. Figure \ref{fig2}(b) is for the explosion reaction. When the temperature reaches a threshold value, the reaction rate increases drastically. Figure \ref{fig2}(c) is for the enzymic catalytic reaction. Neither too high nor too low temperature is of benefit to the activity of enzymes which largely determines the reaction rate. Figure \ref{fig2}(d) is for some kinds of reactions, such as the oxidation of large hydrocarbons. The rise of temperature may has a significant impact on the side reaction inducing a more complicated reaction. Figure \ref{fig2}(e) is for the reaction that $k$ decreases with temperature, which is also a kind of NTC phenomenon \footnote{It should be pointed out that, in many combustion references, Negative Temperature Coefficient (NTC) does not simply represent the fact that the reaction rate decreases with temperature.}. An example for this case is the reaction: $2NO + O_2 \rightarrow 2NO_2$.

\begin{figure}[tbp]
\center\includegraphics*
[bbllx=0pt,bblly=0pt,bburx=180pt,bbury=180pt,angle=0,width=0.7\textwidth]{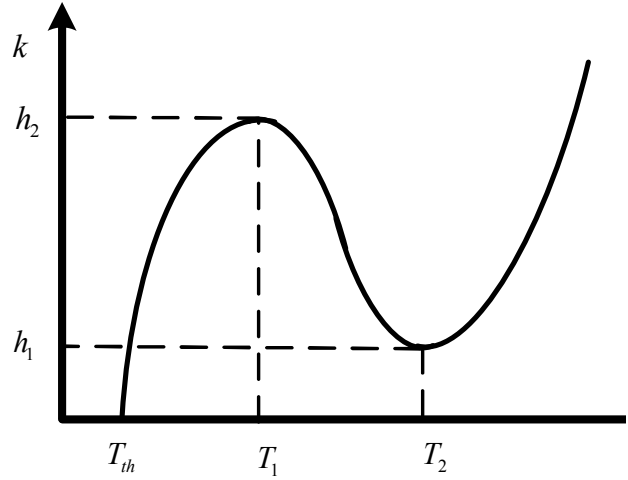}
\caption{Graph of $k(T)$ described by Eq. (\ref{Eq6_k_T}).}\label{fig3}
\end{figure}

To describe the general relation between $k$ and temperature ($T$), a new function $k(T)$ is constructed as follows:
\begin{equation}
k(T)=a+b\int^T_0(t-T_1)(t-T_2)dt \mathtt{,}\label{Eq6_k_T}
\end{equation}
with
\begin{equation}
a=-\frac{-h_2T_1^3+3h_2T_1^2T_2-3h_1T_1T_2^2+h_1T_2^3}{(T_1-T_2)^3} \mathtt{,}\label{Eq7_a_format}
\end{equation}
\begin{equation}
b=-\frac{6(h_1-h_2)}{(T_1-T_2)^3} \mathtt{.}\label{Eq8_b_format}
\end{equation}
where $h_1$ and $h_2$ are the peak and valley values of $k$, respectively. $T_1$ and $T_2$ are temperatures corresponding to $h_1$ and $h_2$, respectively. Its schematic graph is shown in Fig. \ref{fig3}.

To investigate the cases with different kinds of temperature-dependent reaction rates, we need only to adjust the parameters: $h_1$, $h_2$, $T_1$, and  $T_2$, in such a way that the temperature in reaction zone falls into a required range of the curve shown in Fig. \ref{fig3}. In this paper, all the reactant components and the product components are considered as only one kind. The total number of moles do not change during the reaction. So $\lambda$ can be denoted by the concentration of the product. Considering the reaction is irreversible, $\lambda=0$ at the beginning of reaction and $\lambda=1$ at the end of reaction.

\section{Model Verification}
\label{Sec.Verfica}
To demonstrate the validity of the new model, two typical benchmarks of detonation are tested. The first is the  one-dimensional self sustainable stable detonation, and the second is the piston problem. The temporal partial derivative in Eq.(\ref{Eq1}) is solved by first-order forward difference method, and the spatial partial derivative is solved by NND scheme \cite{ODF-book}.
\subsection{Self sustainable stable detonation}
\begin{figure}[tbp]
\center\includegraphics*
[bbllx=0pt,bblly=0pt,bburx=330pt,bbury=510pt,angle=0,width=0.7\textwidth]{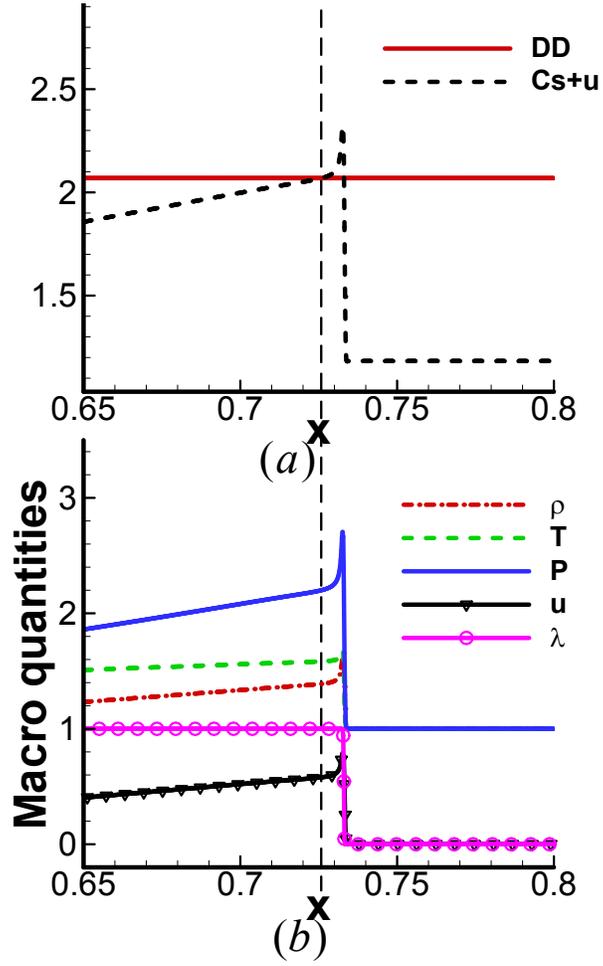}
\caption{Profiles of macroscopic physical quantities at time t=0.35. (a) The position of sound velocity surface, ``$DD$'' indicates the propagation speed of detonation, ``$Cs$'' indicates local sound velocity, and $u$ indicates the fluid velocity behind detonation wave front. (b) Profiles of $\rho$, $u$, $P$, $T$, and $\lambda$. }\label{fig4}
\end{figure}

Consider a rigid tube which is full of premixed combustible gas. At a certain time, a blast start from the left end of the tube. After a while, a self sustainable stable detonation wave will be developed and formed. Initial condition is set as follows:
\begin{equation}
\left\{ \begin{array}{l}
 {(\rho ,u,T,\lambda )_L} = (1.38837,0.57735,1.57856,1) \mathtt{,}\\
 {(\rho ,u,T,\lambda )_R} = (1,0,1,0) \mathtt{.}\\
 \end{array} \right.
\end{equation}
Here the subscripts $L$ and $R$ indicate the left and right side of the domain, respectively. The left boundary is set as a static wall and the right is set to be free flow condition. Other parameters are $\Delta x=\Delta y =2\times 10^{-4}$, $\Delta t=5\times 10^{-6}$, $\tau =2\times 10^{-5}$, $\gamma =1.4$, and the number of grid are $N_x\times N_y=5000\times 1$. Besides, the chemical reaction rate are described by Eq. (\ref{Eq2}) and it has $k=2\times 10^4$, $T_{th}$=1.1, and $Q=1.0$. At the time $t=0.35$, the profiles of hydrodynamic quantities around wave front are shown in Fig. \ref{fig4}(b). The sound velocity surface behind detonation wave is shown in Fig. \ref{fig4}(a). The corresponding values of macroscopic physical quantities in the surface can be obtained from Fig. \ref{fig4}(b) and are shown in Table \ref{Tab1}. In Fig. \ref{fig4}(a), the red solid line indicates the propagation speed of detonation wave, and the black dotted line indicates the sum of velocity of fluid behind detonation and the local sound velocity. The intersection of two lines is the position of sound velocity surface.

Table \ref{Tab1} shows the comparison of DBM simulation results and CJ theoretical values. We can see that relative errors are all less than $0.38\%$, which shows that the new model has a high accuracy in simulating one-dimensional detonation.

\begin{table}[tbp]
\centering  % ±í¾ÓÖÐ
\caption{DBM simulation results compare with CJ theoretical value.}\label{Tab1}
\begin{tabular}{lccc}  % {lccc} ±íʾ¸÷ÁÐÔªËضÔÆ뷽ʽ£¬left-l,right-r,center-c
\hline
 & DBM simulate results & CJ theoretical value & Relative errors(\%)\\ \hline  % \hline ÔÚ´ËÐÐÏÂÃæ»­Ò»ºáÏß
 $D$(Ma) & 1.74870 & 1.74436 & 0.250\\         % \\ ±íʾÖØпªÊ¼Ò»ÐÐ
 $\rho$ & 1.38940 & 1.38837 & 0.074\\         % & ±íʾÁеķָôÏß
 $T$ & 1.58370 & 1.57856 & 0.330\\
 $u$ & 0.57955 & 0.57735 & 0.380\\
 $P$ & 2.19967 & 2.19162 & 0.036\\
$\lambda$ & 1 & 1 & 0 \\ \hline
\end{tabular}

\end{table}

\subsection{Piston problem}
\begin{figure}[tbp]
\center\includegraphics*
[bbllx=0pt,bblly=0pt,bburx=300pt,bbury=280pt,angle=0,width=0.9\textwidth]{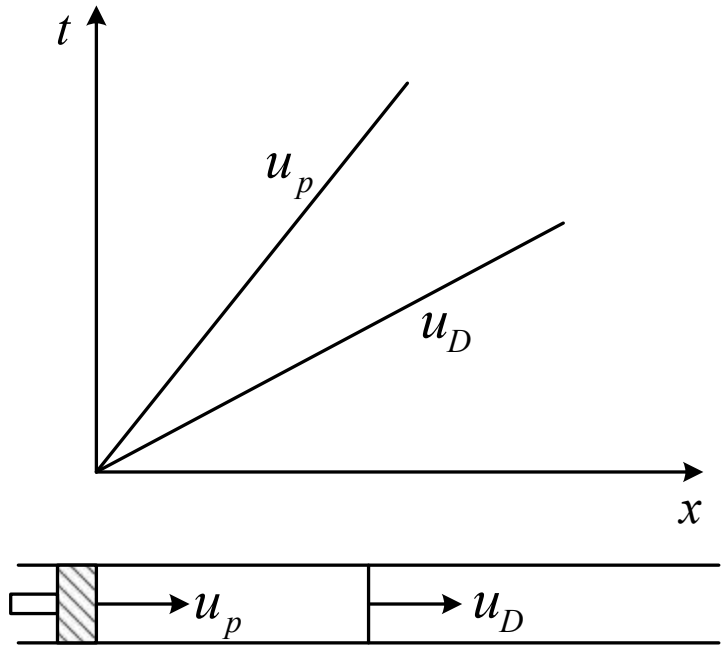}
\caption{Schematic diagram of the piston problem.}\label{fig5}
\end{figure}

As shown in Fig.\ref{fig5}, a detonation wave in rigid tube followed by a piston whose velocity is specified. Due to the influence of the piston, there will be three kinds of different cases according to the relations between the speed of piston and the fluid velocity behind the detonation wave front:

\begin{enumerate}
  \item $u_p > u_{cj}$, where $u_p$ is the speed of piston and $u_{cj}$ is propagation speed of the wave front for CJ detonation. In this case, over-driven detonation will be obtained. Fluid behind the detonation wave is accelerated by piston and compressional waves generate continually until the velocity of the fluid reachs to $u_p$. Then there is a uniform zone between detonation wave front and the piston. The speed of detonation wave is determined by the speed of piston and the detonation energy is provided by both the chemical reaction and the work of piston.
  \item $u_p=u_{cj}$. In this case, CJ detonation will be obtained. The flow speed behind detonation wave equals to the speed of piston, so the piston and detonation wave keep relatively static. Behind the wave there is a uniform zone, the macroscopic physical quantities of fluid behind the wave and the speed of the wave conform well with theoretical value of CJ detonation.
  \item $u_p<u_{cj}$. In this case, CJ detonation followed by rarefaction wave will be obtained. Because the speed of piston is lower than that of fluid behind detonation wave, rarefaction wave develops from the surface of piston. However, the disturbance of rarefaction behind the detonation wave can not catch up with the wave front, so the speed of detonation wave remains unchanged.
\end{enumerate}

\begin{figure}[tbp]
\center\includegraphics*
[bbllx=0pt,bblly=0pt,bburx=570pt,bbury=240pt,angle=0,width=0.9\textwidth]{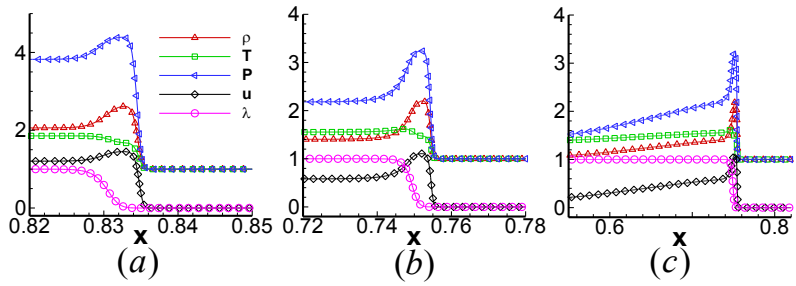}
\caption{Detonation wave charts under different piston speeds. (a) $u_p=1.2$; (b) $u_p=0.57735$; (c) $u_p=0.2$.}\label{fig6}
\end{figure}

 Initial condition set here is the same as those in Fig. \ref{fig4}. Top and bottom boundary conditions are set to be periodic conditions. The left boundary is piston with a constant speed $u_p$ and the right boundary condition is set to be free flow condition. Other parameters are $\Delta x=\Delta y =2\times 10^{-4}$, $\Delta t=5\times 10^{-6}$, $\tau =2\times 10^{-5}$, $\gamma =1.4$, and the number of grid are $N_x\times N_y=5000\times 1$. Chemical reaction conditions are adopted the same as in Ref. \cite{FOP2013-Combustion}. Fig. \ref{fig6} shows spatial distribution of hydrodynamic quantities at the time $t=0.35$. Figures \ref{fig6}(a)-(c) correspond to $u_p=1.2$, $0.57735$, and $0.2$, respectively. From Fig. \ref{fig6} we can see the basic characteristics of detonation including von-Neumann peak, reaction area and rarefaction wave. Table \ref{Tab2} gives the value of macroscopic physical quantities behind detonation wave, it is clear that the detonation type in the case where $u_p > u_{cj}$ is over-driven detonation else it is CJ detonation.

\begin{table}[tbp]
\centering  % ±í¾ÓÖÐ
\caption{DBM simulation results compare with CJ theoretical values}\label{Tab2}
\begin{tabular}{lcccc}  % {lccc} ±íʾ¸÷ÁÐÔªËضÔÆ뷽ʽ£¬left-l,right-r,center-c
\hline
 & $U_p=1.2$ & $U_p=0.57735$ & $U_p=0.2$ & theoretical value\\ \hline  % \hline ÔÚ´ËÐÐÏÂÃæ»­Ò»ºáÏß
 $D$(Ma) & 1.98452 & 1.72415 & 1.72411 & 1.74436\\         % \\ ±íʾÖØпªÊ¼Ò»ÐÐ
 $\rho$ & 2.05194 & 1.39721 & 1.38176 & 1.38837\\         % & ±íʾÁеķָôÏß
 $T$ & 1.86502 & 1.56401 & 1.55704 & 1.57856\\
 $u$ & 1.20390 & 0.58047 & 0.56397 & 0.57735\\
 $P$ & 3.82692 & 2.18525 & 2.15146 & 2.19162\\
 $\lambda$ & 1 & 1 & 1 & 1 \\ \hline
 Type & Over-driven detonation & CJ detonation & CJ detonation & CJ detonation \\  \hline
\end{tabular}
\end{table}

From the simulation results, it is found that over-driven detonation, CJ detonation, and CJ detonation followed with a rarefaction wave area are obtained when $u_p > u_{cj}$, $u_p = u_{cj}$, and $u_p < u_{cj}$, respectively. From Fig. \ref{fig6} and Table \ref{Tab2} we can conclude that the simulation results are well in accordance with the above theoretical ones.

\section{Detonation phenomena under different temperature-dependent reaction rates}
\label{Sec.Simulate}
\subsection{Four kinds of temperature dependent reaction rates}
\begin{figure}[tbp]
\center\includegraphics*
[bbllx=0pt,bblly=0pt,bburx=460pt,bbury=350pt,angle=0,width=0.7\textwidth]{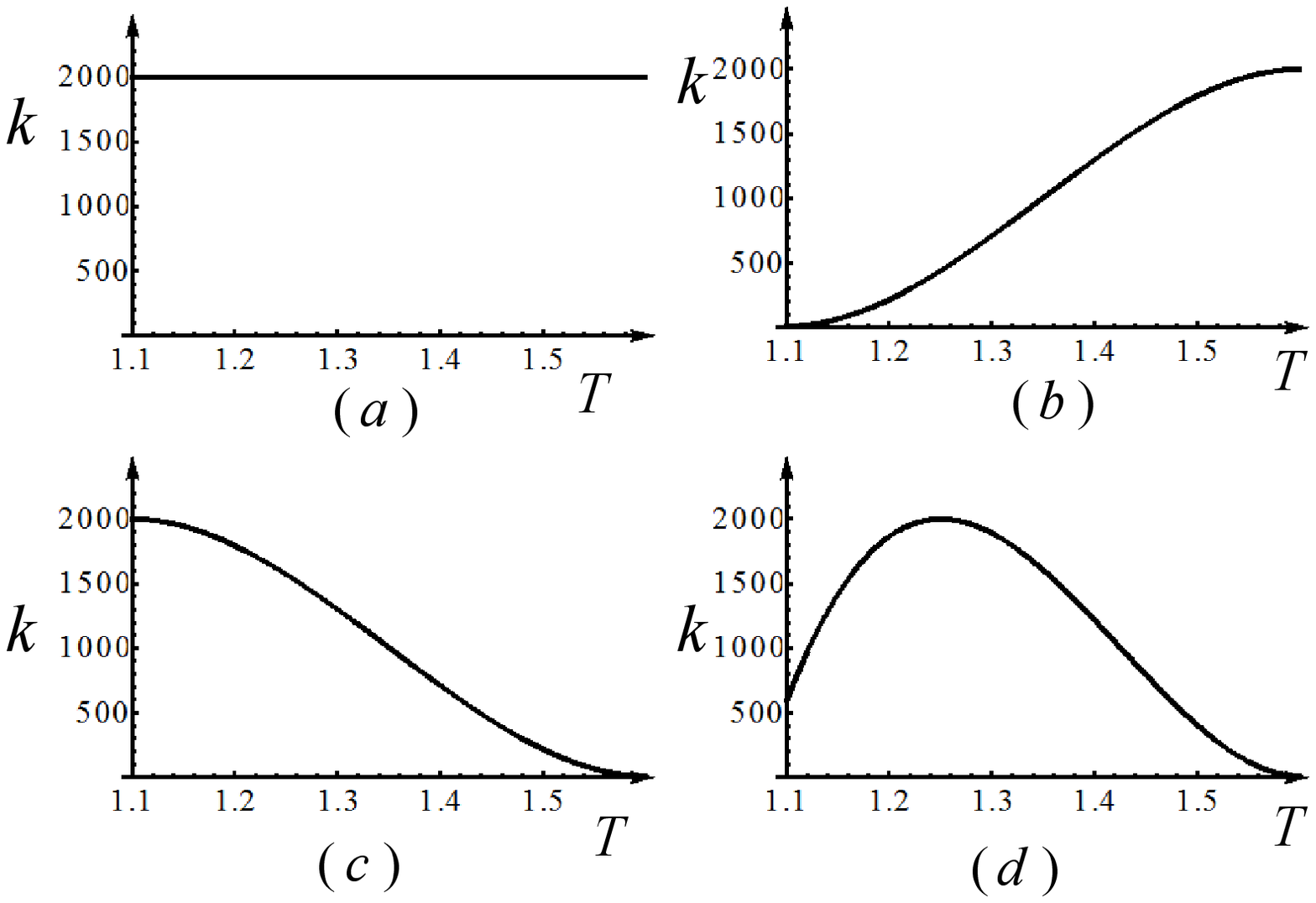}
\caption{Profiles of $k$ with temperature for four different cases. (a)-(d) are corresponding to Case 1-Case 4, respectively.} \label{Fig7}
\end{figure}

In this section, we simulate detonations in four kinds of cases: (i) the reaction rate constant $k$ keeps being a constant, (ii) $k$ increases with increasing temperature, (iii) $k$ decreases with increasing temperature (i.e., NTC), and (iv) $k$ increases firstly and then decreases with increasing temperature. The four kinds of cases are denoted by Case 1, Case 2, Case 3, and Case 4, respectively. The parameters in Eqs. (\ref{Eq6_k_T})-(\ref{Eq8_b_format}) are set as follows:
\begin{itemize}
  \item Case 1:  $T_1=1.1, T_2=1.6, h_1=2000, h_2=2000$;
  \item Case 2:  $T_1=1.1, T_2=1.6, h_1=10, h_2=2000$;
  \item Case 3:  $T_1=1.1, T_2=1.6, h_1=2000, h_2=10$;
  \item Case 4:  $T_1=1.25, T_2=1.6, h_1=2000, h_2=10$.
\end{itemize}
The relations between $k$ and $T$ are shown in Fig. \ref{Fig7}. We will study the difference for the four cases from three aspects in the following section.

\subsection{Hydrodynamic quantities}
\begin{figure}[tbp]
\center\includegraphics*
[bbllx=0pt,bblly=0pt,bburx=550pt,bbury=800pt,angle=0,width=0.9\textwidth]{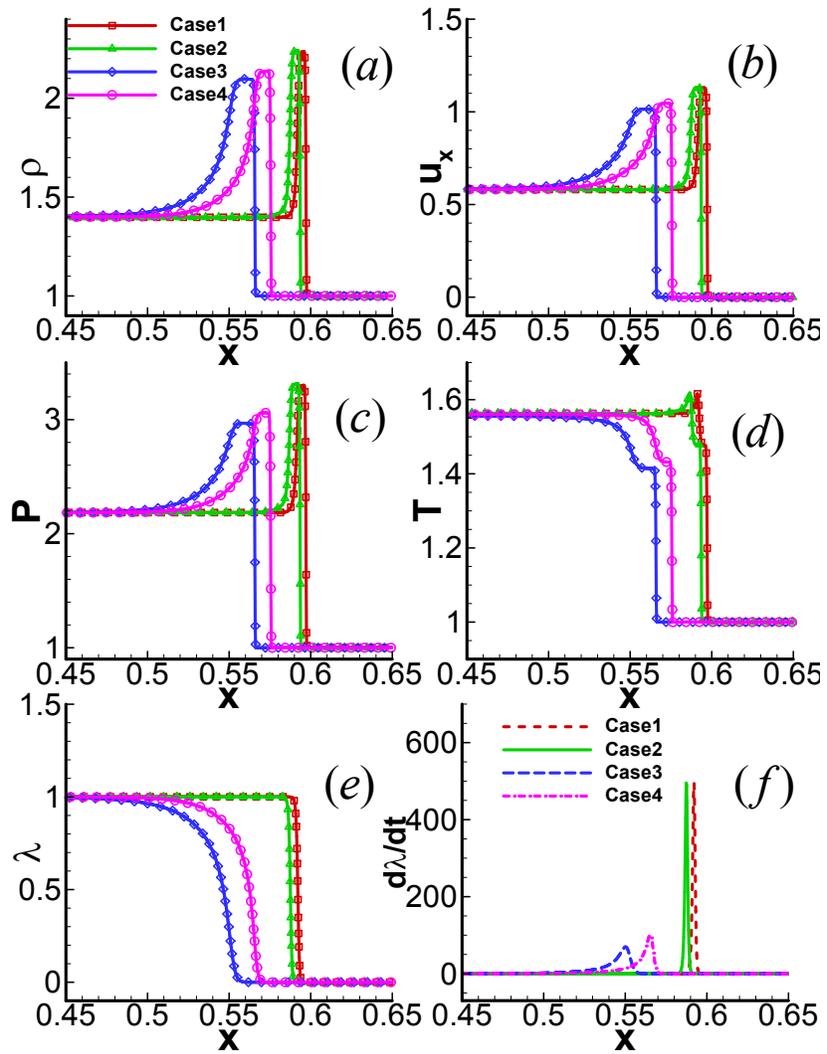}
\caption{Profiles of macroscopic physical quantities.(a)-(f) are for density ($\rho$), x-component of velocity ($u_x$), pressure ($P$), temperature ($T$), mass fraction of product ($\lambda$), and  reaction rate ($d\lambda/dt$) respectively.}  \label{Fig8_Macroscopic quantities}
\end{figure}

The hydrodynamic quantities around detonation wave front are shown in Fig. \ref{Fig8_Macroscopic quantities}. It is found that the profiles for cases 1 and 2 are highly similar to each other, and there is no significant difference in macroscopic physical quantities between the two cases. The reason is that the reaction rate constant determined by temperature for Case 2 approaches to the one for Case 1 in the reaction zone. Besides, it can be found that there is little variation in temperature for Case 2 in the reaction zone from Fig. \ref{Fig8_Macroscopic quantities}(d), so $k$ reaches a stable level at $k=2000$. The reaction rate profiles for Case 1 and Case 2 can be seen from Fig. \ref{Fig8_Macroscopic quantities}(f) and they are almost identical.

\begin{figure}[tbp]
\center\includegraphics*
[bbllx=0pt,bblly=0pt,bburx=330pt,bbury=300pt,angle=0,width=0.7\textwidth]{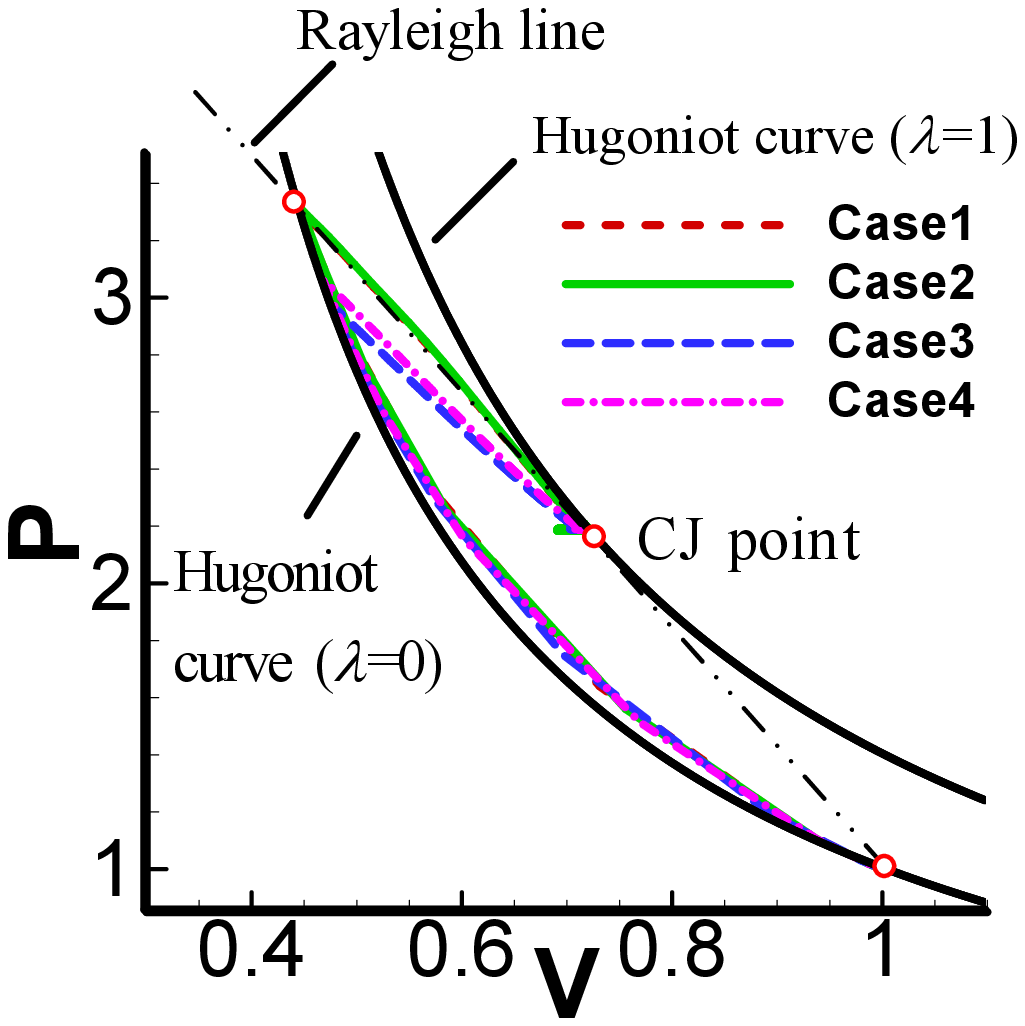}
\caption{P-V phase diagram of the whole calculation domain.}  \label{Fig8_PV}
\end{figure}

From the profiles of $\rho$, $u_x$, and $P$ in Fig. \ref{Fig8_Macroscopic quantities}(a), (b), (c), it can be seen that there is a lower peak value and a wider wave structure for Case 3 and Case 4 than those for Case 1 and Case 2. The change is more obvious for Case 3 than Case 4. In fact, Case 3 has a complete NTC while Case 4 contains a partial NTC. So the characteristic in Case 4 should be somewhere in between Case 2 and Case 3. From the above analysis, it can be concluded that the role of NTC is to lower the peak value and broaden the wave structure of $\rho, u_x, P$. The reason can be found in Fig. \ref{Fig8_Macroscopic quantities}(e) and (f). From Fig. \ref{Fig8_Macroscopic quantities}(e), it can be seen than chemical time are much longer for case 3 and 4 than cases 1 and 2, and it also can be seen that the distribution of reaction rates for cases 3 and 4 have a lower amplitude and wider regions than cases 1 and 2 from Fig. \ref{Fig8_Macroscopic quantities}(f). That is why cases 3 and 4 has a wider wave structure. Further, a wider wave structure leads a stronger rarefaction behind the detonation wave front, which is the reason why the peaks of $\rho$, $u_x$, and $P$ are lower for cases 3 and 4 than cases 1 and 2.

From the profiles of $T$ in Fig. \ref{Fig8_Macroscopic quantities}(d), it can be seen that temperatures increase rapidly at first, then reach to a peak value and decrease toward a steady-state value for cases 1 and 2. However, the temperatures rise slowly and have no peak value during the whole process for cases 3 and 4. We can conclude that the role of NTC is to smooth the variation in temperature and vanish the peak value of $T$.

In our model, we assume that the chemical reaction will start as soon as the temperature reaches a threshold value ($T_{th}$). It is possible for the chemical reaction to start and consequently release heat in the shocking stage. Before the von-Neumann peak,
    there are two kinds of mechanisms, shocking and reaction, to increase the temperature.
    Behind the von-Neumann peak,
    the chemical reaction is to increase the temperature, while the rarefaction effect is to decrease the temperature. The behaviour of the temperature is determined by the competition of the two mechanisms. So, the temperature may show non-monotonic behaviour.

 For cases 1 and 2, the chemical reaction is very quick, it has nearly finished around the von-Neumann pressure peak, consequently there is nearly no more heat released after the von-Neumann pressure peak. That is the reason why we observe a decrease of temperature. For case 3 or 4, the situation is significantly different. The NTC makes the reaction more slowly. After the von-Neumann pressure peak, the chemical reaction is still in process and the reaction heat is still continuously added into the system. At the same time, the rarefaction effect here is weaker than in cases 1 and 2. Thus, we do not observe the decrease of temperature in cases 3 and 4. For cases 3 and 4, the decrease of temperature may result in a more violent detonation, as shown in Figs. 7(e)-(f), which is responsible for that the temperature peak disappears for cases 3 and 4.

From the profiles of $\lambda$ and $d\lambda/dt$ in Fig. \ref{Fig8_Macroscopic quantities}(e) and (f), it can be seen that the chemical reactions complete in a very short time and the reactions are very violent for case 1 and 2. However, for cases 3 and 4, the reaction zones have an obvious extension and the reaction rate amplitudes is much lower than those for cases 1 and 2. We can conclude that another role of NTC is to lower reaction rate and prolong the reaction time.

Figure \ref{Fig8_PV} shows the P-V diagram. From this figure we can see that detonation process develop as follows: firstly, the pressure and density increase along the Hugoniot curve with $\lambda=0$ till the pressure and density reach to their peak values, then they begin to make a transition from the peak values to their stable state, i.e., CJ point which is in the Hugoniot curve with $\lambda=1$, and chemical reaction completes within the transition period. It can be found that the four cases have similar P-V diagrams, but the peak value for Case 3 is lower than Case 4, and they are both lower than those for cases 1 and 2. However, the final state behind the wave structure for the four cases are almost the same. From the Rayleigh relation we can conclude that the speed of detonation wave structure should be the same value for the four cases. So, it can be concluded that the NTC lower the peak value of density and pressure but has no effect on the detonation process and the speed of wave structure.

%%The propagation speeds of detonation wave are shown in Fig. \ref{Fig8_Macroscopic quantities}(d). It is obtained by tracking the positions of pressure peak at each time step and shifting the curves of the four cases to the same starting point. From the $x$-$t$ phase diagram, it is obvious that the wave speeds for Case 1 and Case 2 are faster than those for Case 3 and Case 4. The reason is that detonation wave is driven by chemical reaction after the wavefront. Since the role of NTC is lower the reaction rate,  it is intelligible for the lowering of corresponding wave speed.

\subsection{Non-equilibrium effects}
\label{Sec.non-equil}
Considering that all cases in this paper are for one-dimensional issues, only the $\Delta^*_{2,xx}$ and $\Delta^*_{3,1,x}$ of non-equilibrium tensor, defined in Eqs. (\ref{Eq13_Delta2*}) and (\ref{Eq14_Delta31*}) are used in this section.

\begin{figure}[tbp]
\center\includegraphics*
[bbllx=0pt,bblly=0pt,bburx=510pt,bbury=240pt,angle=0,width=0.9\textwidth]{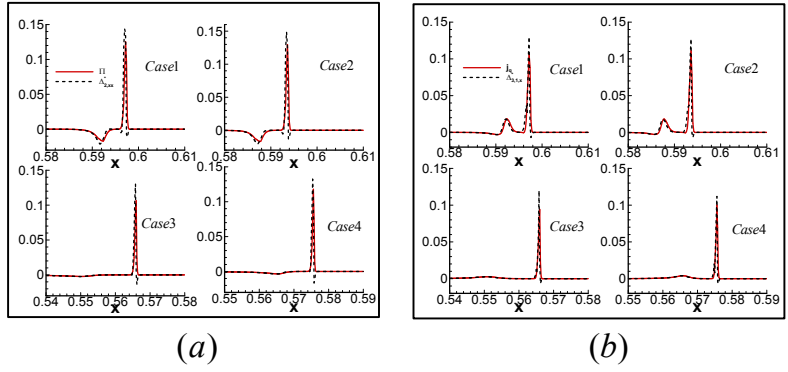}
\caption{Comparisons of viscous stress, heat flux and non-equilibrium quantities. (a)$\Pi_{xx}$ and $\Delta^*_{2,xx}$. (b)$j_{q,x}$ and $\Delta^*_{3,1,x}$.} \label{Fig9_1_non-equilibrium fig}
\end{figure}

Firstly, the viscous stress ($\Pi_{xx}$) and heat flux ($j_{q,x}$) are compared with the non-equilibrium quantities, respectively, in Fig. \ref{Fig9_1_non-equilibrium fig}(a) and (b).
 It can be seen that, for the four cases $\Pi_{xx}$ are in accordance well with $\Delta^*_{2,xx}$ in the region where the system are in or near equilibrium state, but there are observable deviations between the two terms in the region where the system have significant deviations from equilibrium state. There are the same characteristics between $j_{q,x}$ and $\Delta^*_{3,1,x}$.

In section \ref{Sec.Models}, we have known that $\Delta^*_{2,xx}$ and $\Delta^*_{3,1,x}$ in the Eq. (\ref{Eq16_momentum conservation equation}) and (\ref{Eq17_energy conservation equation}) are derived directly from Boltzmann equation while $\Pi_{xx}$ and $j_{q,x}$ are derived based on the approximation, $f \approx f^{0} + f^{1}$. It is a reasonable approximation when the system is near the equilibrium state, but it would be inaccurate when the system deviates far from equilibrium sate. This means that NS may be imperfect to describe detonation, especially in the region near the detonation wave front where non-equlibrium effect is remarkable.

\begin{figure}[tbp]
\center\includegraphics*
[bbllx=0pt,bblly=0pt,bburx=610pt,bbury=280pt,angle=0,width=0.9\textwidth]{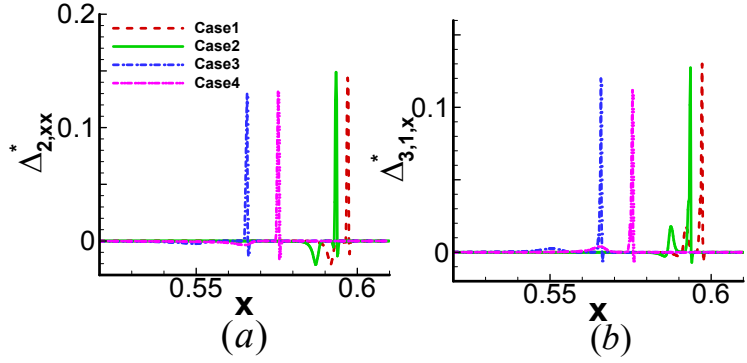}
\caption{Profiles of non-equilibrium effects for four cases. (a)$\Delta^*_{2,xx}$. (b)$\Delta^*_{3,1,x}$.} \label{Fig9_2_non-equilibrium fig}
\end{figure}

The profiles of $\Delta^*_{2,xx}$ and $\Delta^*_{3,1,x}$ for four cases are shown together in Fig. \ref{Fig9_2_non-equilibrium fig}. It can be seen that the amplitudes of $\Delta^*_{2,xx}$ and $\Delta^*_{3,1,x}$ in reaction zones for cases 3 and 4 are lower than those for cases 1 and 2, and the non-equilibrium regions for cases 3 and 4 are wider than those for cases 1 and 2. Furthermore, the amplitudes of $\Delta^*_{2,xx}$ and $\Delta^*_{3,1,x}$ around the detonation wave front for cases 3 and 4 are also lower than those for cases 1 and 2. Because the deviation from thermodynamical equilibrium state in reaction zone is mainly caused by chemical reaction rate, while the NTC lower amplitude of the reaction rate and extend the reaction zone. So the NTC lower the strength of non-equilibrium effect and wider the non-equilibrium zone.

\subsection{Entropy production}
The three terms of entropy production in Eq. (\ref{Eq21_entropy production}) can be denoted by $\sigma_1$, $\sigma_2$, and $\sigma_3$, respectively. The global entropy production in the whole calculation domain can be denoted by $\Delta s_1$, $\Delta s_2$, and $\Delta s_3$, respectively. Then it has
\begin{subequations}

\begin{equation}
\sigma_1=\mathbf{\Delta}_{3,1}^*\cdot \bigtriangledown(\frac{1}{T}) \mathtt{,} \label{Eq22_a}
\end{equation}
\begin{equation}
\sigma_2=-\frac{1}{T}\mathbf{\Delta}_2^*:\bigtriangledown\mathbf{u}  \mathtt{,} \label{Eq22_b}
\end{equation}
\begin{equation}
\sigma_3=\rho\frac{Q}{T}F(\lambda)  \mathtt{.} \label{Eq22_c}
\end{equation}
\end{subequations}

and
\begin{equation}
\Delta s_i=\int\sigma_i dV {\kern 20pt} (i=1,2,3) \label{Eq23}   \mathtt{.}
\end{equation}

\begin{figure}[tbp]
\center\includegraphics*
[bbllx=0pt,bblly=0pt,bburx=580pt,bbury=400pt,angle=0,width=0.9\textwidth]{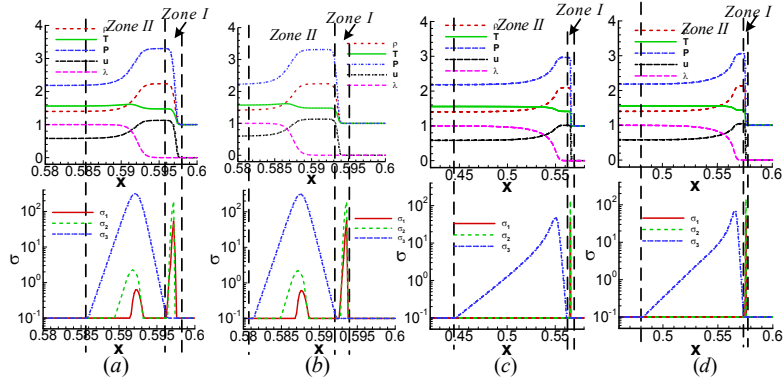}
\caption{Three kinds of profiles of entropy productions. (a)-(d) are for Case 1 - Case 4, respectively.} \label{Fig10_entropy production fig}
\end{figure}

\begin{figure}[tbp]
\center\includegraphics*
[bbllx=0pt,bblly=0pt,bburx=510pt,bbury=340pt,angle=0,width=0.9\textwidth]{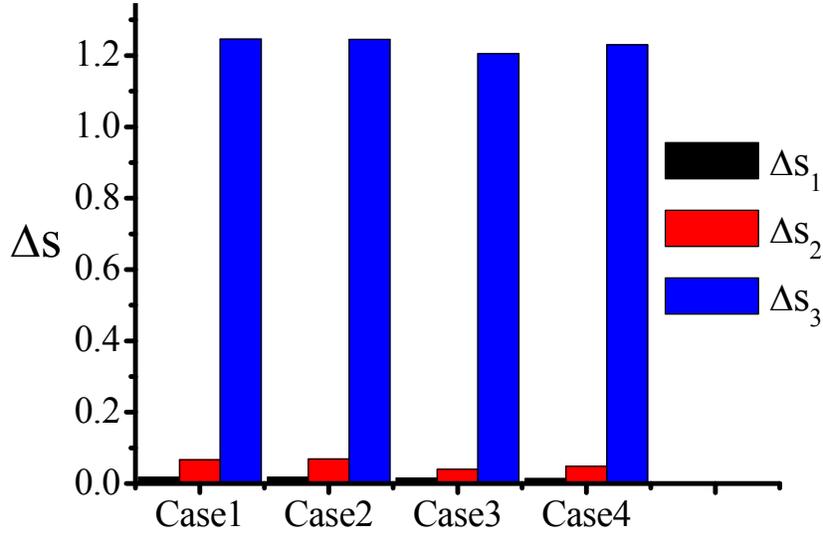}
\caption{Three kinds of profiles of global entropy productions for four cases. } \label{Fig11_global entropy production fig}
\end{figure}

The local entropy productions($\sigma_i,i=1,2,3$) and global entropy productions
($\Delta s_i, i=1,2,3$) of four cases are shown in Fig. \ref{Fig10_entropy production fig} and Fig. \ref{Fig11_global entropy production fig}, respectively.

From Fig. \ref{Fig10_entropy production fig}, it can be seen that there are two regions with entropy production in the process of detonation. The first one is in near the wavefront where the entropy production mainly caused by NOEF ($\sigma_1$) and NOMF ($\sigma_2$), while the second one happens in the chemical reaction zone where the entropy production mainly caused by chemical reaction ($\sigma_3$). The magnitude of $\sigma_2$ is higher than $\sigma_1$ and they are both much lower than $\sigma_3$. For cases 1 and 2, $\sigma_3$ have concentrated distributions in reaction zone. While for cases 3 and 4, $\sigma_3$ have relatively sparse distributions and the peak values are lower than those for cases 1 and 2. Besides, this change for Case 3 is more obvious than Case 4. The entropy productions, $\sigma_1$ and $\sigma_2$, for cases 3 and 4 are not obvious.
From Fig. \ref{Fig11_global entropy production fig}, it can be seen that the global entropy production is mainly caused by chemical reaction. It has the relation that $\Delta s_3$ $>>$ $\Delta s_2$ $>$ $\Delta s_1$. Because it is a high-Mach propagation process for detonation, the role of NOMF is more important than that of NOEF in entropy production. Besides, with the increasing of Mach number, the entropy production caused by NOMF becomes more remarkable.
Comparing the four cases, we can see that the three global entropy productions of Case 3 are all smaller than those of Case 4 and they both are smaller than those of Case 1 and Case 2.
As for the effect of NTC on $\Delta s_2$ and $\Delta s_1$, because NTC reduces reaction rate and the intensity of reaction, lowers the impact strength, drives the detonation to be closer to the isentropic process, consequently it lowers the value of $\Delta s_2$ and $\Delta s_1$. While as for the effect on $\Delta s_3$, the detonation with NTC has a lower reaction rate which increases the extent of quasi-static, so the value of $\Delta s_3$ is lower than in the cases without NTC.

\section{Conclusions}
\label{Sec.Conclu}

To study the combustion system, both the kinetic and hydrodynamic models are revisited.
 A new version of hydrodynamic model is presented.  The relations between non-equilibrium quantities and entropy productions are established. Based on the DBM with a new reaction rate model, four kinds of detonations with different temperature-dependent reaction rates are simulated. The behaviours of the four cases are comparatively studied through three aspects: hydrodynamic quantities, non-equilibrium effects, and entropy productions.

From the side of hydrodynamic quantities, it is concluded that the role of NTC is to lower the peak of density, pressure, and velocity in reaction zone, to broaden the reaction zone, and to lower the instantaneous strength of reaction. It may also vanish the peak of temperature. The reason is that reaction rate determined by temperature in reaction zone is comparatively lower for the cases containing NTC than those not. Besides, $k$ would continue to decrease with the rise of temperature, which results in a further decrease of reaction rate.

From the side of non-equilibrium, comparisons are made between viscous stress and NOMF, heat flux and NOEF.
The numerical results of
viscous stress (heat flux) have a good agreement with those of
 NOMF (NOEF) near equilibrium state. Around sharp interfaces, the values of NOMF (NOEF) deviate reasonably from those of viscous stress (heat flux).
Besides, the role of NTC is to lower amplitude of non-equilibrium effect and broaden the non-equilibrium zone in reaction zone.

From the side of entropy production, it is found that the portion of global entropy production caused by the reaction is much larger than those by NOMF and NOEF. Entropy production caused by NOMF is much larger than that caused by NOEF (i.e., $\sigma_2 > \sigma_1$ and $\Delta s_2 > \Delta s_1$). Besides, NTC deconcentrates the distribution of entropy production caused by chemical reaction and lower the global entropy productions caused by chemical reaction, NOMF and NOEF. This means that detonation cases containing NTC are more likely to follow an isentropic and quasi-static process.

\section*{Acknowledgements}
The authors would like to sincerely thanks Drs. Yanbiao Gan, Huilin Lai, Zhipeng Liu for helpful discussions. AX and GZ acknowledge support of the Foundation of LCP, National Natural Science Foundation of China [under Grant No. 11475028].

\end{document}